\documentclass[prl,reprint,twocolumn,superscriptaddress]{revtex4-1}
\usepackage[utf8]{inputenc}
\usepackage[T1]{fontenc}
\usepackage{textcomp}
\usepackage{graphicx}
\usepackage{epstopdf}
\usepackage{longtable}
\usepackage{dcolumn}
\usepackage{bm}
\usepackage{amsmath}
\usepackage{xcolor}
\usepackage{times}
\newcommand{\namno}{NaMn$_7$O$_{12}$}

\newcommand{\camno}{CaMn$_7$O$_{12}$}
\newcommand{\ymno}{YMn$_7$O$_{12}$}
\newcommand{\lamno}{LaMn$_7$O$_{12}$}
\newcommand{\amno}{$A$Mn$_7$O$_{12}$}

\begin{document}
\title{Unconventional magnetic ferroelectricity in the quadruple perovskite \namno}

\author{Vinícius Pascotto Gastaldo}
\email[Corresponding Author. Current address: Universidade Federal de Mato Grosso do Sul, Av. Costa e Silva, s/nº Campo Grande - MS, Brazil. ~E-mail: ]{vinicius.gastaldo@ufms.br}
\affiliation{Departamento de F\'isica, Universidade Federal de S\~ao Carlos, Rodovia Washington Lu\'is, km 235, S\~ao Carlos - S\~ao Paulo, Brazil}
\affiliation{IMPMC-Sorbonne Universit\'e and CNRS, 4, place Jussieu, 75005 Paris, France}

\author{Yannick Klein}
\affiliation{IMPMC-Sorbonne Universit\'e and CNRS, 4, place Jussieu, 75005 Paris, France}

\author{Beno\^it Baptiste}
\affiliation{IMPMC-Sorbonne Universit\'e and CNRS, 4, place Jussieu, 75005 Paris, France}

\author{Riccardo Cabassi}
\author{Edmondo Gilioli}
\affiliation{Istituto IMEM-CNR, Area delle Scienze, 43100 Parma, Italy}

\author{Andrea Gauzzi}
\affiliation{IMPMC-Sorbonne Universit\'e and CNRS, 4, place Jussieu, 75005 Paris, France}

\author{Adilson J. A. de Oliveira}
\affiliation{Departamento de F\'isica, Universidade Federal de S\~ao Carlos, Rodovia Washington Lu\'is, km 235, S\~ao Carlos - S\~ao Paulo, Brazil}

\date{\today}

\begin{abstract}
By means of magnetic, specific heat and pyroelectric measurements, we report on magnetic ferroelectricity in the quadruple perovskite \namno, characterized by a canted antiferromagnetic (AFM) CE structure. Surprisingly, ferroelectricity is concomitant to a dramatic broadening of the magnetic hysteresis loop, well below the AFM ordering temperature. This unconventional behavior shows that the formation of ferroelectric domains is induced by the symmetric exchange interaction in the local scale, \textit{e.g.} at magnetic domain boundaries or defects. The value of electric polarization, $P = 0.027 \mu$C cm$^{-2}$, measured in polycrystalline samples is comparatively large as compared to other magnetic multiferroics, suggesting that the above scenario is promising indeed for the rational design of practical multiferroic materials.
\end{abstract}

\maketitle

\section{Introduction}

Manganese oxides, such as simple perovskites $RE$MnO$_3$ ($RE$ = rare earth) with non-collinear E-type antiferromagnetic (AFM) structure \cite{kim03} and quadruple perovskites \amno\ ($A$=Ca, La) with canted AFM structure \cite{zha18,joh18,lamnomultif,ver19}, have attracted a great deal of interest for they display enhanced magnetic ferroelectricity (MF). Among these compounds, \lamno, first reported by Bochu \textit{et al.} \cite{boc74}, exhibits record values of remanent polarization at saturation as high as 0.56 $\mu$C cm$^{-2}$ in polycrystalline samples \cite{lamnomultif}. Even higher values 1-10 $\mu$C cm$^{-2}$ are expected in single crystals or epitaxial films due to the absence of depolarizing fields, which may enable the realization of novel applications like fast, nonvolatile and low-energy-consumption memories \cite{eereinstein, fiebig, multifmedia}.

A prerequisite for these important developments is to establish clear relationships between the structural and electronic properties of the above compounds and the enhancement of polarization observed experimentally. In this respect, a number of points remain open or controversial. A first point concerns the magnetically induced noncentrosymmetric distortions of the lattice expected in any MF compound that remain elusive experimentally or limited to tiny structural modulations \cite{wal11}. A second point is the controversial contribution of the antisymmetric or symmetric exchange interactions to the electric polarization \cite{ser06,pic07,lu12} and the possible role of charge order \cite{kho06,zha13}. The existence of a pronounced canting of the AFM structure in \camno\ and \lamno, caused by a strong Dzyaloshinskii-Moriya interaction, led various authors to propose that the symmetric exchange interaction should be the dominant mechanism of ferroelectricity in the \amno\ system \cite{lu12,zha13,lamnomultif}. However, the complex structural modulations concomitant to the magnetic order observed in the most studied compound \camno\ \cite{per12} prevents to establish clear relationships between ferroelectric and magnetic properties.

In order to elucidate the above points, which would be an important step towards a rational design of multiferroic materials, in the present work, we investigate the occurrence of magnetic ferroelectricity in \namno\ \cite{mar73} that shares with \camno\ and \lamno\ a similar quadruple perovskite (QP) crystal structure (see Fig. \ref{MC}) and a strongly canted AFM CE-type structure \cite{namnomag}. Our purpose is to single out the structural and electronic properties that make the QP system favorable to host large magnetic ferroelectricity.

\section{The Quadruple Perovskite \namno}

The QP structure, first reported by Marezio \textit{et al.} \cite{mar73} and described by the general formula $(AA'_3)B_4$O$_{12}$, differs from the simple perovskite structure $AB$O$_3$ for the presence of a Jahn-Teller active $A'$ site driving a large tilt of the corner-sharing $B$O$_6$ octahedra, which leads to a competition between ferromagnetic (FM) and antiferromagnetic (AFM) superexchange interactions between neighboring $A'$ or $B$ ions \cite{ver17}. The presence of two distinct $A$ and $A'$ sites enables to obtain mixed-valence properties without introducing disorder or chemical inhomogeneities inherent to chemically-substituted $AB$O$_3$ compounds like manganites \cite{sal01} or nickelates \cite{hot06}. These unique characteristics may be the key to explain the variety of unique charge, orbital and spin ordering phenomena observed in QP compounds \cite{boc79,gauzzinat,lon16} as well as the aforementioned remarkable enhancement of magnetic ferroelectricity in the restricted QP \amno\ family, where both $A'$ and $B$ sites are occupied by Mn \cite{camnopol2,camnopol1,lamnomultif,ver19}.  
 
The $A$=Na compound is interesting owing to a simple cubic $Im\bar3$ structure at room temperature \cite{mar73}. By lowering temperature, \namno\ undergoes an almost full Mn$^{3+}$/Mn$^{4+}$ charge order (CO) transition of the octahedral Mn $B$ sites at $T_{CO}=176$ K \cite{gauzzinat} accompanied by a monoclinic $I2/m$ distortion. The CO order is precursor of a canted AFM CE-type order of the $B$ sites at $T_{N,B}$=125 K. At $T_{N,A'}$=90 K, the Mn $A'$ sites also order to form a AFM G-type structure. A feature relevant to the present work is a history-dependent exchange bias-like displacement of the magnetic hysteresis loop, attributed to uncompensated spins in the AFM structure \cite{namnomag}.

\section{Experimental}

We studied both single- and poly-crystalline samples synthesized under high pressure at 6 GPa and 1000 K, as described in detail elsewhere \cite{sint}. An x-ray diffraction characterization of the latter samples indicate that their purity is better than 95\%. Magnetization, $M$, measurements as a function of temperature in the 2 - 300 K range at fields up to $H=7$ T were performed on both types of samples in a commercial Quantum Design VSM-SQUID MPMS3 magnetometer. For a precise determination of the coercive fields and of the remanent magnetization, $M$ vs $H$ measurements were carried out at constant temperature after having quenched the superconducting magnet. This enabled a preliminary calibration using a Pd sample as standard and ensured a precise determination of the field by eliminating the spurious effect of the remnant magnetic flux trapped in the magnet. The specific heat of polycrystalline samples was measured using a commercial Quantum Design PPMS apparatus in the 2 - 300 K range at fields up to 9 T using a $2\tau$ relaxation method. We prepared the same polycrystalline samples in the form of thin disks suitable for pyroelectric current measurements performed using a Keithley 617 electrometer and a home-made sample holder installed inside the MPMS3 cryostat. The disks were mechanically polished to mirror finish, which prevents the accumulation of trapped charges at the sample surface that may give a spurious contribution to the pyroelectric signal, as demonstrated previously \cite{ngo15}. The samples were first poled at 50 K for 30 minutes under an electric field of 2.4 kV/cm, then field-cooled down to 15 K and short-circuited for 10 minutes to release any trapped charges \cite{jac17}. Finally, the pyroelectric current was measured upon warming the sample at a 5 K/min rate up to room temperature. This procedure was repeated at different poling fields or warming rates. As complementary characterization of the polycrystalline samples, we measured the AC dielectric constant from impedance measurements using a commercial HP4824A LCR meter and the liquid helium cryostat of a commercial MPMS SQUID magnetometer. For these measurements, polycrystalline samples were lined with metallized mica sheets to avoid the Maxwell-Wagner effect \cite{wag14} at the contact-sample interface and an open-short correction was applied to correct for residual impedance of the set-up wiring, as described elsewhere \cite{dielec}.

\section{Magnetization and specific heat measurements}

\begin{figure}[ht]
	\centering
	\includegraphics[width=\linewidth]{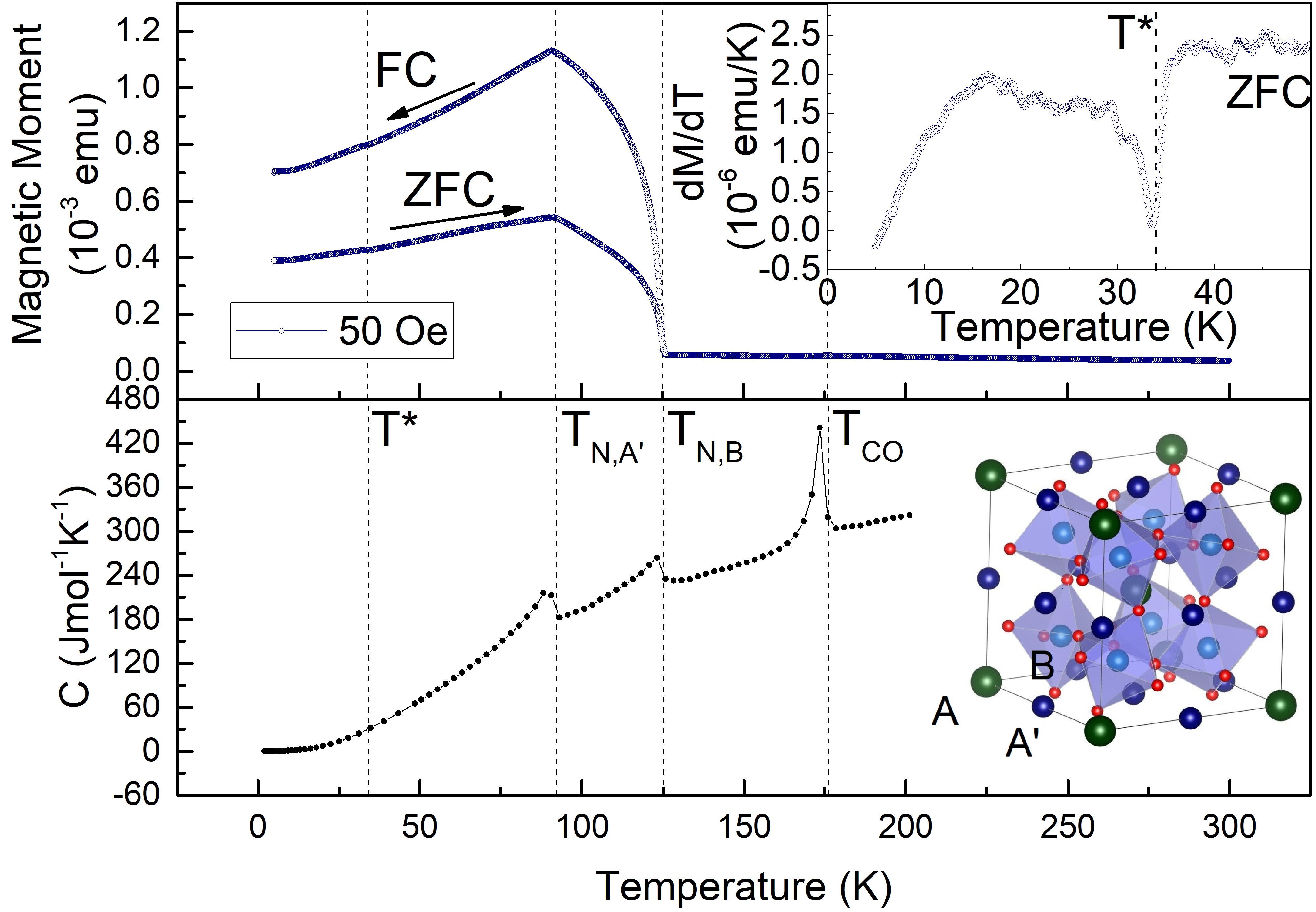}
	\caption{Temperature dependence of the magnetization $M(T)$ measured in low field at 50 Oe (top) and of the zero-field specific heat $C(T)$ (bottom) of a polycrystalline \namno\ sample. Broken vertical lines indicate the CO and AFM ordering temperatures of the $A'$ and $B$ sublattices and the anomaly in the $M(T)$ curve at $T^{\ast}=34$ K. This anomaly is seen as a pronounced peak in the $dM/dT$ curve of the upper inset. Bottom inset: the pseudo-cubic unit cell of quadruple perovskites $(AA'_3)B_4$O$_{12}$. In \namno, $A$=Na and $A',B$=Mn. Oxygen atoms are represented by red spheres.}
	\label{MC}
\end{figure}

Fig. \ref{MC} shows the temperature dependence of the low field ($H$=50 Oe) magnetization and of the zero-field specific heat of a polycrystalline \namno\ sample. In agreement with previous reports \cite{gauzzinat,namnomag}, both curves display the characteristic signatures of the CO transition at $T_{CO}=176$ K and of the two AFM orderings at $T_{N,A'}= 92$ K and $T_{N,B} = 125$ K. Surprisingly, the present high-resolution magnetization data unveil a further anomaly at $T^{\ast}=34$ K, not reported before. The anomaly consists of a clear peak in the $dM/dT$ curve (see inset of Fig. \ref{MC}). In the $C(T)$ data, the anomaly is less visible and manifests itself as a broad hump in the derivative curve, $dC/dT$ (data not shown). The absence of a jump in the $C(T)$ curve points at a subtle change in the ground state that rules out the scenario of a second-order phase transition, in agreement with previous neutron diffraction data \cite{gauzzinat} showing no additional magnetic Bragg peaks below $T^{\ast}$. We recall that similar subtle anomalies in the temperature-dependence of the specific heat have been previously reported in improper ferroelectrics like Gd$_2$(MoO$_4$)$_3$ or Tb$_2$(MoO$_4$)$_3$ \cite{improp} and hydrogen-based ferroelectric lawsonite \cite{impropexp}.

\begin{figure}[ht]
	\centering
	\includegraphics[width=\linewidth]{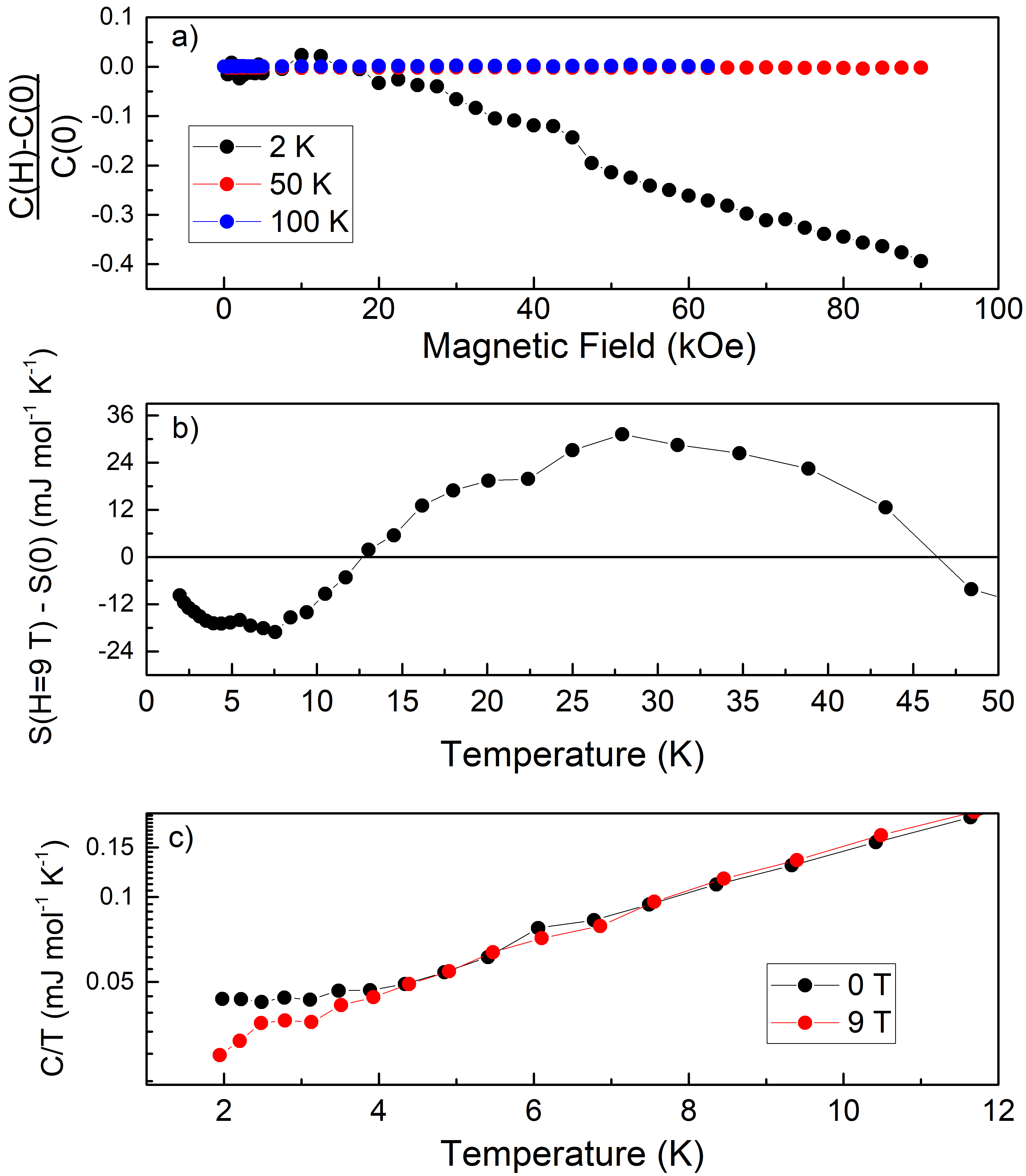}
	\caption{(a): Relative variation of the specific heat with field at 2, 50 and 100 K. (b): Temperature dependence of the entropy difference $\Delta S = S(T,H=9 {\rm T})-S(T,H=0)$ showing a sizable field-induced reduction (increase) of magnetic entropy below (above) $\approx 15$ K. The entropy $S$ has been computed by integrating the $H=0$ and $H=9$ T $C(T)/T$ curves of panel (c) according to eq. (1).}
	\label{anomaly}
\end{figure} 

In order to unveil the origin of the above anomaly in the $M(T)$ curve, we have studied the field dependence of the specific heat of a polycrystalline \namno\ sample at various temperatures. Indeed, the anomaly is expected to affect the field dependence of the isochoric specific heat $C_V(H)$, according to Maxwell's relation $(\partial M / \partial T)_{V,H} = (\partial S / \partial H)_{V,T}$, where the entropy $S$ is obtained from $C_V$ as:

\begin{equation}
S(T,H)=\int_0^T \frac{C_V(T,H)}{T}dT
\end{equation}

In fact, our specific heat data are isobaric, but the isobaric volume change of \namno\ is negligible in the low-temperature range considered, as reported previously \cite{gauzzinat}. The result of our study, summarized in Fig. \ref{anomaly}a, confirms the above expectation. At low temperature, $T=2$ K, we find a large linear decrease of $C(H)$ with $H$ reaching 40\% at 9 T. The effect of field vanishes at high temperatures above 50 K, as seen in Fig. \ref{anomaly}a. A physical interpretation of the above result is provided by plotting the difference $\Delta S = S(T,H=9 {\rm T})-S(T,H=0)$ as a function of temperature (see Fig. \ref{anomaly}b). Note that the field causes a sizable reduction of entropy $\approx$20-30 mJ mol$^{-1}$K$^{-1}$ below $\approx$15 K followed by a comparable increase above this temperature. This transfer of magnetic entropy from lower to higher temperatures is reminiscent of a Schottky anomaly caused by the Zeeman splitting of two or more energy levels separated by a characteristic energy $\sim k_BT$ with $T \sim 15$ K, corresponding to magnetic defects or different spin configurations in the unit cell \cite{lue99,gop66}. Considering an entropy contribution of $k_B \ln 2$ for a $S=1/2$ spin, the magnitude of the entropy transfer corresponds to one defect every $\sim 200$ formula units, in agreement with the magnitude of the remanent magnetization \cite{namnomag}.    

\begin{figure}[ht]
	\centering
	\includegraphics[width=\linewidth]{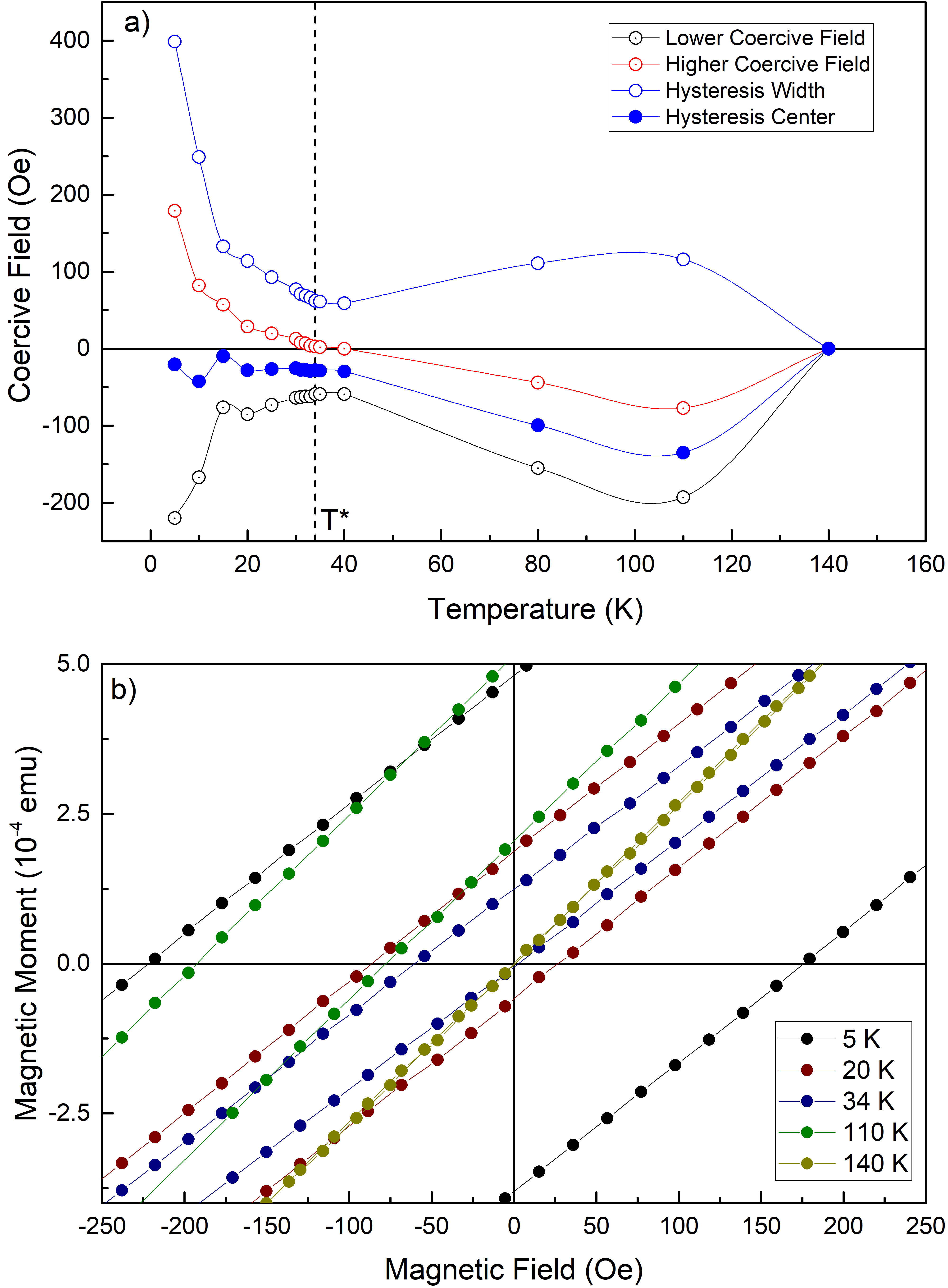}
	\caption{(a) Temperature evolution of the magnetic hysteresis loop position and width across the anomaly at $T^{\ast}$ in a \namno\ polycrystalline sample. The lower (upper) coercive field is defined as the value at which the loop crosses the negative (positive) field axis, while the hysteresis width is the difference between upper and lower coercive fields. (b) Plot of the loops at representative temperatures.}
	\label{hyst}
\end{figure}

The above field dependence points at a change of magnetic entropy at $T^{\ast}$ that preserves the long-range order of the AFM CE structure. We have attempted to investigate this change by means of field-dependent magnetization measurements at various temperatures above and below $T^{\ast}$. The results are summarized in Fig. \ref{hyst}. In agreement with a previous study \cite{namnomag}, we have found a remanent magnetization attributed to uncompensated AFM moments. The present systematic study of the loops as a function of temperature further unveils a sizable and steady increase of the loop width below $T^{\ast}$, accompanied by a shift of the loop center towards the origin at $H=0$. Above $T^{\ast}$, the loop width increases moderately again before disappearing in the paramagnetic phase above $T_{N,B}$. These trends are reproducible from sample to sample in both poly- and single-crystalline samples, which indicates that the effects are intrinsic and not associated with intergranular or exchange bias effects at grain boundaries. We conclude that the loop broadening reflects either an increase of the canted component of the AFM CE structure or an enhancement of the magnetization at the domain boundary. The existence of an exchange bias in the loops would be consistent with the above picture of magnetic defects. Indeed, a defect in the AFM structure generates an uncompensated spin and thus an internal field.
 
\section{Pyroelectric current and dielectric measurements}

\begin{figure}[ht]
	\centering
	\includegraphics[width=\linewidth]{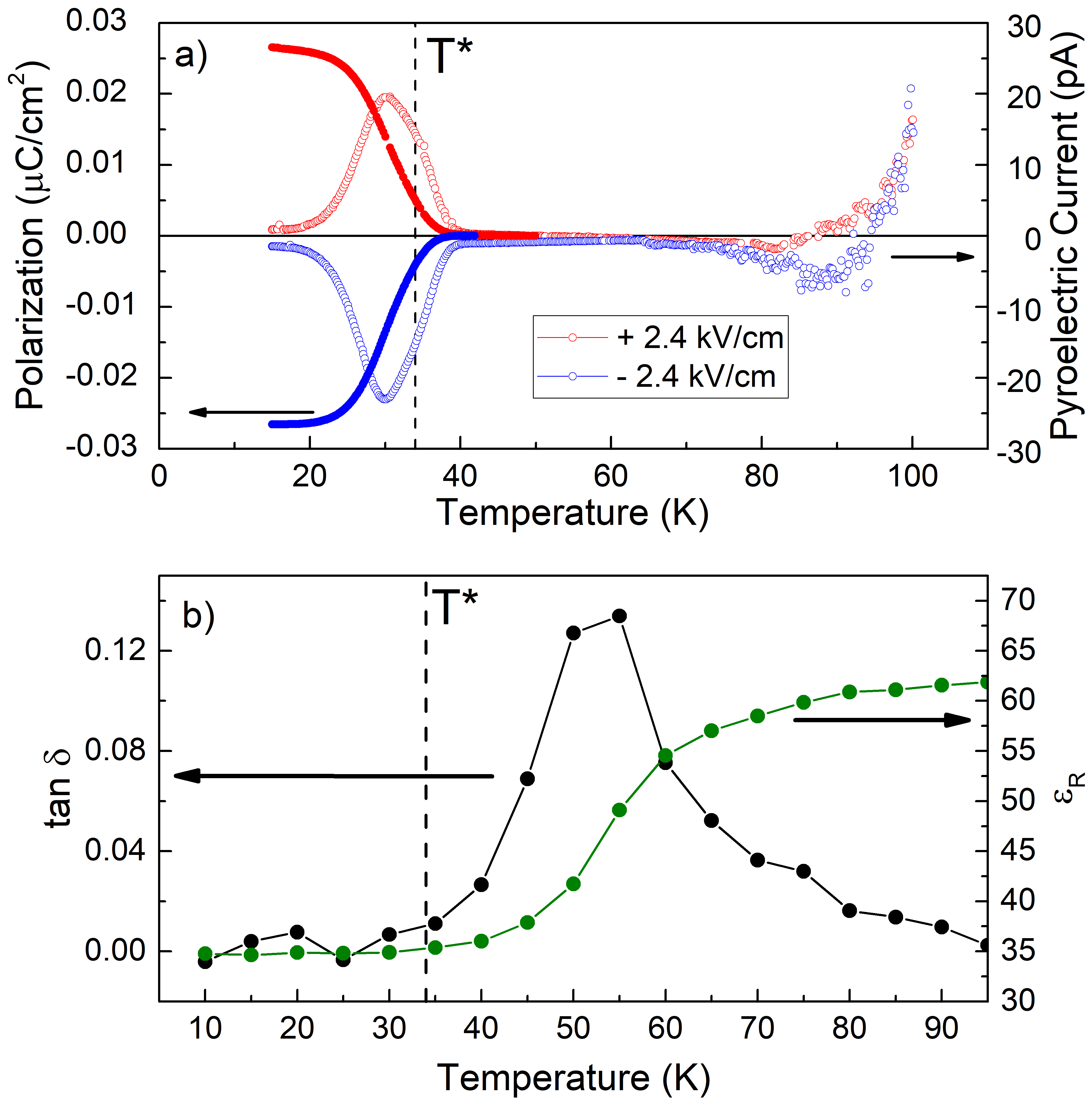}
	\caption{(a) Pyrocurrent and remanent polarization curves of the same \namno\ polycrystalline sample used for the magnetic measurements of Fig. \ref{hyst}. Red and blue curves refer to poling fields of opposite polarity, $\pm$2.4 kV/cm, respectively. The broken line indicates the characteristic temperature $T^{\ast}$ of the magnetic anomaly in Fig. \ref{MC}. (b) Temperature dependence of the relative dielectric constant and of the loss factor for a similar polycrystalline sample (from \cite{dielec}).}
	\label{pyro}
\end{figure}

Following the aforementioned reports on magnetic ferroelectricity in the \amno\ system, we investigated whether \namno\ as well displays a similar behavior by means of pyroelectric current measurements on the same polycrystalline sample used for the magnetic measurements. Remarkably, we did find a strong 10 K broad pyrocurrent peak centered near $T^{\ast}$, \textit{i.e.} well below the AFM ordering temperature (see Fig. \ref{pyro}). As expected for a ferroelectric transition, the sign of the pyrocurrent changes with the sign of the poling field. Owing to the careful preparation of the sample surface and of the short circuit imposed to the opposite sides of the disk at low temperature prior to the measurements (see above), we rule out any spurious contribution of thermally activated currents associated with trapped charges, as reported previously \cite{ngo15,artefato}. A spurious contribution appears only above $\approx$100 K as a current growing exponentially with temperature. Note that this current does not change appreciably with the sign of the poling field, contrary to the pyrocurrent peak at $T^{\ast}$. As customary, the remanent electric polarization is obtained by integrating the peak area, which yields $P=0.027 \mu$C cm$^{-2}$ (see Fig. \ref{pyro}). In order to compare this value with previous results obtained in the related compounds \camno\ and \lamno, further measurements as a function of poling field are required to determine the $P$ value at saturation. Here, we limit ourselves to note that, in \camno\ polycrystalline samples, a slightly lower value $P=0.024 \mu$C cm$^{-2}$ has been reported for a 40\% higher poling field of 3.5 kV/cm \cite{camnopol2}. We conclude that, to the best of our knowledge, \namno\ displays the second best ferroelectric performances among magnetic ferroelectrics after \lamno. The same measurements repeated in magnetic fields up to 7 T did not produce any measurable change in the pyroelectric current. The scenario of ferroelectric transition that we propose here is supported by previous dielectric constant measurements carried out on a similar polycrystalline \namno\ sample at 1 kHz \cite{dielec}. The data in the inset of Fig. \ref{pyro} show a pronounced peak of the loss factor accompanied by a drop of the relative dielectric constant just above $T^{\ast}$. Both features indicate that the dielectric properties of the system drastically change below $T^{\ast}$. The absence of a peak in the dielectric constant, characteristic of conventional ferroelectrics, is consistent with the absence of a specific heat jump at $T^{\ast}$ or, equivalently, the absence of a second-order phase transition. The above phenomenology reflects the unconventional magnetic ferroelectric behavior of \namno, namely the onset of ferroelectricity does not occur at the magnetic ordering temperature, as it is usually the case of magnetic ferroelectrics, including the related QP compounds \camno\ and \lamno\ mentioned before. Instead, in \namno, ferroelectricity is concomitant to a magnetic anomaly occurring well below this temperature. A similar feature has been reported recently in \ymno, for which a scenario of ferroelectricity induced by a spin glass behavior has been proposed \cite{ver19}.

\section{Discussion and conclusions}

We should finally attempt to explain the anomalous interplay between magnetic and ferroelectric behavior observed. To do so, we recall the two different scenarios proposed before to account for the magnetic hysteresis results of Fig. \ref{hyst}. According to the first scenario that invokes an enhancement of the canted component of the AFM structure at $T^{\ast}$, the symmetric exchange interaction associated with this component would be the driving mechanism of ferroelectricity, as already proposed for both \camno \cite{lu12} and \lamno \cite{lamnomultif}. The second scenario of net magnetization at the domain boundary is consistent with a theoretical model of AFM magnetoelectrics predicting the formation of a net magnetization at the AFM domain boundary \cite{domain}. If this was the case, the broadening of the hysteresis loops in Fig. \ref{hyst} would be the consequence - instead of the cause - of ferroelectricity. In both scenarios, the present results suggests that the formation of ferroelectric domains is driven by the symmetric exchange interaction in the local scale, \textit{e.g.} at magnetic domain boundaries or defects.

In summary, we have given evidence of a large magnetic ferroelectricity in the mixed-valence quadruple perovskite \namno. Remarkably, the onset of ferroelectricity occurs well below the AFM ordering temperature, at a characteristic temperature $T^{\ast}$ where the the magnetic hysteresis loops display an anomalous broadening. The ferroelectric phase is therefore unconventional, for it is not concomitant to the appearance of any secondary order parameter as in the case of improper ferroelectrics and couples with magnetic properties in ways so far not reported. To account for the observation of ferroelectricity at $T^{\ast}$, we propose two alternative scenarios involving either the enhancement of the canted component of the AFM structure or the formation of a net magnetization at the AFM domain boundary.

To discriminate between the two scenarios, high-resolution neutron diffraction studies would be required to investigate the details of the magnetic structure and to detect the changes of this structure at $T^{\ast}$. These studies may unveil the long-sought noncentrosymmetric distortions of the lattice expected in magnetic ferroelectrics. Further pyroelectric current or ferroelectric hysteresis measurements on single-crystals would be needed to determine the orientation of the electric polarization and thus to establish a quantitative model of magnetic ferroelectricity. In both scenarios, the present results give evidence of a link between canted antiferromagnetism driven by the Dzyaloshinskii-Moriya interaction and large magnetic ferroelectricity, thus providing a hint for the rational design of practical multiferroic materials. 

\section{Acknowledgements}
\begin{acknowledgements}
The authors acknowledge Im\`ene Est\`eve and David Hrabovsky for their valuable assistance in sample preparation and in the specific heat measurements and gratefully acknowledge the Brazilian and French foundations FAPESP, CAPES and COFECUB for financial support through the grants FAPESP 2012/21171-5, 2013/07296-2, 2013/27097-4, 2017/24995-2 and 2015/21206-1 and the CAPES-COFECUB Project No. Ph 878-17 (88887.130195/2017-01). The authors also acknowledge the support of UFMS.
\end{acknowledgements}

\bibliography{biblio}

\end{document}